\documentclass[aps,prd,preprintnumbers,showpacs,nofootinbib,floatfix,amssymb,preprint,a4paper]{revtex4-1}

\usepackage{amsfonts,amsmath,units,wasysym,epsfig,graphicx,verbatim,color,subfigure,graphicx,bm,mathrsfs,lipsum,hyperref}

\begin{document}

\newcommand{\IUCAA}{Inter-University Centre for Astronomy and
  Astrophysics, Post Bag 4, Ganeshkhind, Pune 411 007, India}

\newcommand{\WSU}{$^*$Department of Physics \& Astronomy, Washington State University,
1245 Webster, Pullman, WA 99164-2814, U.S.A \\}

\title{Quasi-normal Modes of Static Spherically Symmetric Black Holes in $f(R)$ Theory}
\author{Sayak Datta}\email{skdatta@iucaa.in} 
\affiliation{\IUCAA}

\author{Sukanta Bose}\email{sukanta@iucaa.in}
\affiliation{\IUCAA}
\affiliation{\WSU}

\date{\today}

\begin{abstract}

We study the quasi-normal modes (QNMs) of static, spherically symmetric black holes in $f(R)$ theories. We show how these modes in theories with non-trivial $f(R)$ are fundamentally different from those in General Relativity. In the special case of $f(R) = \alpha R^2$ theories, it has been recently argued that iso-spectrality between scalar and vector modes breaks down. Here, we show that such a break down is quite general across all $f(R)$ theories, as long as they satisfy $f''(0)/(1+f''(0)) \neq 0$, where a prime denotes derivative of the function with respect to its argument. We specifically discuss the origin of the breaking of isospectrality. We also show that along with this breaking the QNMs receive a correction that arises when $f''(0)/(1+f'(0)) \neq 0$ owing to the inhomogeneous term that it introduces in the mode equation. We discuss how these differences affect the ``ringdown" phase of binary black hole mergers and the possibility of constraining $f(R)$ models with gravitational-wave observations. We also find that even though the iso-spectrality is broken in $f(R)$ theories, in general,  nevertheless in the corresponding scalar-tensor theories in the Einstein frame it is unbroken.
\end{abstract}

\preprint{LIGO-P1900111}

\maketitle

\section{Introduction}

Recent observations of gravitational wave (GW) signals in LIGO and Virgo from compact object mergers have opened up a new chapter in the history of physics~\cite{LIGO detection 1,LIGO detection2,LIGO detection3,LIGO detection4}. So far these detectors have already observed several binary black hole mergers and a binary neutron star merger~\cite{First BNS observation}. The latter observation has put strong constraints on the equation of state of the matter inside neutron stars~\cite{LSC constrain on EOS,Soumi De 2018,Tianqi Zhao 2018,Eemeli Annala 2018,Radicel 2018}. It has also provided a strong bound on the graviton mass \cite{Abott 2017 graviton mass} and the deviation of the velocity of GWs from that of the light \cite{First BNS observation,Baker 2017}. As a result it has already helped us understand not only the mergers themselves but also fundamental aspects of the nature of gravity~\cite{Baker 2017,Ezquiaga 2017,Abott 2016, Lucas 2016, Lucas 2017, Konoplya 2016}. This scenario is expected to get even more interesting when LISA is launched in several years from now since the space detector will give us the opportunity to test gravity in a different frequency band~\cite{LISA 1,LISA 2,LISA 3,LISA 4,LISA 5}.

To date general relativity (GR) has been very successful in observational tests of its predictions
(see, e.g., Ref.~\cite{Will book,will 2014}). 
However, 
certain aspects of GR, e.g., black hole and cosmological spacetime singularities, have invited proposals for
higher-curvature correction terms to the Einstein-Hilbert action~\cite{Stelle 1977,Stelle 1978,Shahid 1990,Brandenberger 1992,Brandenberger 1993,Brandenberger et al. 1993,Trodden 1993, Nojiri 2011, Nojiri 2017} in attempts toward resolving these pathologies. These corrections 
come in various forms,
such as functions of Ricci scalars and different combinations of the contractions of the Riemann tensor. 
These suggestions have gained importance due to other reasons as well. It has been shown that several actions with correction terms can drive inflation and others have been successful in explaining the late time acceleration of the universe~\cite{Starobinsky 1980,Hu,Nojiri 2003, Cognola 2008}. Such actions have also arisen from the low-energy limit of quantum corrections or String Theory \cite{Birrell 1982,Buchbinder 1992,Vilkovisky 1992} and in Loop Quantum Gravity \cite{Zhang 2011,olmo 2009}. These results have initiated extensive research in such alternative theories of gravity in the last few decades. Several studies have already investigated modifications in QNMs for other types of corrections (quadratic in curvature)~\cite{Zinhailo 2018, Konoplya 2019}. Now the availability of GW observations helps us examine if these theories can be subjected to useful tests.

As has been shown in Numerical Relativity, when the two black holes in a binary system merge, a distorted black hole is created. This remnant radiates GWs as it settles down into a Kerr black hole
\cite{Regge 1957,Zerilli 1970,Newman 1962,Vishu 1970,Chandra}. 
Asymptotically, these GWs can be expressed as superpositions of damped sinusoidal modes, termed as quasi normal modes (QNMs). These QNMs in GR depend only on a couple of parameters characterizing the black hole, namely, its mass and spin, for astrophysical black holes. This is a manifestation of the No-Hair theorem \cite{Nollert 1999,Kokkotas 1999,Konoplaya 2011}. Therefore,   observation of these QNMs in the future is an anticipated testing ground for the No-Hair theorem and alternative theories of gravity.

It is well known that in GR there are two modes that contribute to GW observables. These are the odd (vector) and even (scalar) modes, and they share the net emitted gravitational energy equally. It has been argued in Ref.~\cite{Shanki, Soham 2018} that in the case of the $(R+\alpha R^2)$ action, where $R$ is the Ricci scalar, QNMs emitted from black holes will not have the same spectrum as in GR. 
In this work, we show that such a break down in {\em iso-spectrality}
is quite general across all $f(R)$ theories, as long as they satisfy $f''(0)/(1+f'(0)) \neq 0$, where a prime denotes derivative of the function with respect to its argument. 
We specifically discuss the origin of the breaking of isospectrality. We also show that along with this breaking
the QNMs receive a correction that arises when $f''(0)/(1+f'(0) \neq 0$ owing to the inhomogeneous term that it introduces in the mode equation.
We also discuss the resulting structural modification of the QNMs. 

In an important piece of work, Tattersall et al.~\cite{Tattersall 2018_1, Tattersall 2018_2} demonstrated that in Horndeski theory of gravity the QNM equation acquires a source term, which results in the breaking of iso-spectrality. Going by these results alone, one might expect that when an $f(R)$ theory is related to an STT through a conformal transformation, and both theories have the Schwarzschild solution, then iso-spectrality should break in that $f(R)$ theory as well for QNMS in that solution.
Interestingly, and perhaps somewhat counterintuitively, we show here that when an $f(R)$ theory and its conformally related STT both have the Schwarzschild metric as solution, and the value of the conformal factor relating these respective solutions is unity, then isopectrality is not broken for the Schwarzschild QNMs in STT even though it is broken for the Schwarzschild QNMs in $f(R)$.

In Sec.~\ref{General Result} general results of an arbitrary $f(R)$ theory and the modified Einstein equations in such theories are discussed. In Sec.~\ref{Formalism} we present the perturbation equations in the context of GR around the Schwarzschild metric. In Sec.~\ref{Perturbation in general f(R)} we discuss the perturbation of the modified Einstein equation in general $f(R)$ theories. In Sec.~\ref{perturbation modes} we deduce the different modes and the final perturbation equations in $f(R)$ theories. Then in Sec.~\ref{solution for qnm} we deduce the changes in the QNMs in comparison to GR. Finally, in Sec.~\ref{conclusion and discussion} we summarize our results. Throughout the work we have adopted the $\{-,+,+,+\}$ signature and set $G = c =1$.

\section{General equations for $f(R)$ theories} 
\label{General Result}

As mentioned above, 
here we focus our attention on $f(R)$ theories of gravity. Our primary motivation for doing so is that they offer resolutions for certain limitations in the standard cosmological model and the fact that $f(R)$ terms arise as correction terms from Loop Quantum Gravity. 
Another reason is that these theories are not affected by the Ostrogradsky instability~\cite{Woodard 2007}. Among all the $f(R)$ theories we only focus on those that allow $R =0$ solutions, for the obvious reason that such solutions exist in GR and are observationally relevant. We discuss this aspect in more detail later.

Let us begin by considering the gravity model that has a Lagrangian of the following form:
\begin{equation}
\label{action}
S_f = \frac{1}{2\kappa^2}\int d^4x~\sqrt{-g}~f(R),
\end{equation}
where $\kappa^2 = 8\pi$.
Varying this action with respect to the metric $g_{\mu\nu}$ yields the equation of motion~\cite{Myung},
\begin{equation}
\label{eom of f(R)}
R_{\mu\nu} f'(R)-\frac{1}{2}f(R) g_{\mu\nu} + (g_{\mu\nu}\Box - \nabla_{\mu}\nabla_{\nu})f'(R) = 0.
\end{equation}
It is known that Eq.~(\ref{eom of f(R)}) has a constant curvature solution, $R = \bar{R}$. In that case, the above equation takes the form 
\begin{equation}
\bar{R}_{\mu\nu}f'(\bar{R}) - \frac{1}{2}g_{\mu\nu}f(\bar{R}) = 0.
\end{equation}
Taking its trace gives
\begin{equation}
\label{Constraint on f(R) due to zero curvature}
\bar{R}f'(\bar{R}) -2 f(\bar{R}) = 0.
\end{equation}
Since our objective here is to study perturbations of the Schwarzschild black hole, we focus only on such models of $f(R)$ that have $\bar{R} = 0$ as a root of Eq.~(\ref{Constraint on f(R) due to zero curvature}). It is straightforward to verify that GR is but one example of such theories.

Since our main goal is to look for deviations from GR, we will separately track the GR part of the $f(R)$ action, namely, the $R$ term. Therefore, we will express the Lagrangian as
\begin{equation}
f(R) = R + \varphi (R)\,,
\end{equation}
where $\varphi (R)$ denotes terms in the gravity action beyond GR. Henceforth, we rename $\varphi (R)$ as $f(R)$. Thus, our working Lagrangian becomes $R+f(R)$. 

\section{QNMS in General Relativity} 
\label{Formalism}

For the purpose of our calculations we adopt the gauge invariant formalism discussed in Refs.~\cite{Ishibasi 2011,kodama 2000,kodama 2003,Gerlach 1979}. Our focus remains the Schwarzschild spacetime. In the case of a spherically symmetric spacetime, it is possible to split the manifold into an orbit space and a unit sphere. We will be focusing on the four-dimensional non-spinning black hole background. In that case the manifold splits into a two-dimensional orbit space and a two-sphere. We will use the co-ordinates $x^a$ for the orbit space and $z^A$ for the two-sphere. The covariant derivatives on the orbit space and the two-sphere are represented by $D_a$ and $D_A$ respectively. The d'Alembertian operators on the orbit space and the two-sphere are represented by $\tilde{\Box}$ and $\hat{\Box}$, respectively.

With proper choice of co-ordinates it is possible to cast the metric in the following form,
\begin{equation}
\label{eq:background}
d\bar{s}^2 = -g(r) dt^2+\frac{1}{g(r)}dr^2+r^2d\Omega^2 ,
\end{equation}
where the overbar represents quantities of the unperturbed background, $g(r) = 1-\frac{2M}{r}$, and $d\Omega^2$ represents the metric on the unit two-sphere.
The metric perturbation on the background can be expressed as follows~\cite{Ishibasi 2011,kodama 2000,kodama 2003,Gerlach 1979}:
\begin{equation}
ds_p^2 = h_{ab}dx^a dx^b +2 h_{aB}dx^a dz^B +h_{AB}dz^Adz^B,
\end{equation}
where $ds_p^2$ represents the perturbed part of the line element.

To simplify the problem, the metric functions are usually expanded in spherical harmonic basis. After linearizing the Einstein equations, they can be cast in terms of the co-ordinates of the orbit space alone. This problem has been well studied. We follow the formalism defined by Kodama, Ishibashi and Seto (KIS). 
With proper identification of the scalar and vector mode it is possible to show that for each multipole $l\geq 2$ they satisfy the following equations~\cite{Ishibasi 2011,kodama 2000,kodama 2003}:
\begin{widetext}
\begin{equation}
\label{ GR qnm eqn}
\begin{split}
&\frac{d^2\Phi^{GR}_S}{dr_*^2}+ (\omega^2-V_S)\Phi^{GR}_S = 0,\\
&\frac{d^2\Phi^{GR}_V}{dr_*^2}+ (\omega^2-V_V)\Phi^{GR}_V = 0.
\end{split}
\end{equation}
\end{widetext}

\begin{equation}
V_{S/V} = W^2 \, \mp \frac{d W}{dr_*} -\frac{\lambda^2(\lambda + 1)^2}{9M^2},\,\,\,\,\,\,\,\,\,\,\,\,\,\,\,\,W = \frac{6M(2M-r)}{r^2(6M+2\lambda r)}-\frac{\lambda(\lambda + 1)}{3M},
\end{equation}
where $\lambda = \frac{(l-1)(l+2)}{2}$ and the time dependence has been taken to be $e^{-i\omega t}$. The subscripts $S$ and $V$ represent the scalar mode~(Zerilli mode) and the vector mode~(Regge-Wheeler mode), respectively. Moreover, ``GR'' in the superscript distinguishes these modes from those in $f(R)$. 

These QNMs have been studied extensively in the literature. It has been shown that the transmission and the reflection coefficients of the $V_S$ and $V_V$ are equal. They make equal contributions to GWs asymptotically. It has also been demonstrated that they share the same frequency spectrum~\cite{Chandra}. We will show below that this does not hold, in general, in $f(R)$ theories.

\section{Perturbation in general $f(R)$} 
\label{Perturbation in general f(R)}

Our main goal is to investigate the QNM structure of the Schwarzschild black hole. Therefore, we take the background metric to be that given in Eq.~(\ref{eq:background}).
Now the perturbed metric can be written as
\begin{equation}\label{Perturbed metric}
g_{\mu\nu} = \bar{g}_{\mu\nu} + h_{\mu\nu}\,,
\end{equation}
where $\bar{g}_{\mu\nu}$ and $h_{\mu\nu}$ represent the background metric and its perturbation, respectively.
Using the perturbed metric from Eq.~(\ref{Perturbed metric}) in the equation of motion Eq.~(\ref{eom of f(R)}), we find the equation for the perturbation as follows,
\begin{equation}
\begin{split}
\label{perturbation eqn 1}
(1+f^{\prime}(\bar{R}))\delta R_{\mu\nu}(h)&-(\bar{R}+f(\bar{R}))\frac{h_{\mu\nu}}{2}+f^{\prime\prime}\bigg[\bar{g}_{\mu\nu}\bar{\Box} +\frac{\bar{R}}{4}\bar{g}_{\mu\nu}\\
&-\bar{\nabla}_{\mu}\bar{\nabla}_{\nu}-\frac{(1+f^{\prime}(\bar{R}))}{2f^{\prime\prime}(\bar{R})}\bar{g}_{\mu\nu}\bigg]\delta R(h) = 0,
\end{split}
\end{equation}
where $\delta R_{\mu\nu}$ and $\delta R$ are the perturbed Ricci tensor and Ricci scalar, respectively. Equation~(\ref{perturbation eqn 1}) can be re-expressed as
\begin{equation}
\delta G^{eff}_{\mu\nu} = \kappa^2 \delta T^{eff}_{\mu\nu},
\end{equation}
where
\begin{equation}
\begin{split}
&\delta G^{eff}_{\mu\nu}=\bigg[\delta \bar{R}_{\mu\nu}-\frac{1}{2}\bar{g}_{\mu\nu}\delta R\bigg]-\frac{(\bar{R}+f(\bar{R}))}{(1+f^{\prime}(\bar{R}))}\frac{h_{\mu\nu}}{2},\\
&\delta T^{eff}_{\mu\nu}=-\frac{f^{\prime\prime}}{\kappa^2(1+f^{\prime}(\bar{R}))}\bigg[\bar{g}_{\mu\nu}\bar{\Box} +\frac{\bar{R}}{4}\bar{g}_{\mu\nu}-\bar{\nabla}_{\mu}\bar{\nabla}_{\nu}\bigg]\delta R(h).
\end{split}
\end{equation}
Taking the trace of Eq.~(\ref{perturbation eqn 1}) it can be shown that
\begin{equation}
\label{massive scalar eqn 1}
3f^{\prime\prime}\bar{\Box}\delta R-(1+f^{\prime}(\bar{R}))\delta R-(\bar{R}+f(\bar{R}))\frac{h}{2} +f^{\prime\prime}\bar{R}\delta R = 0\,,
\end{equation}
where $h$ is the trace of the perturbation. As we show more explicitly below this equation contains a massive scalar degree of freedom that is generically present in $f(R)$ theories, unlike general relativity. Further details are available in Refs.~\cite{Capozziello 2008,Nishizawa 2009,Berry}. This equation is equivalent to the equation of a massive scalar field with a source term that depends on the perturbation.

\subsection{QNMs in $f(R)$} \label{perturbation modes}

In the present work our main goal is to find the perturbed equation of Schwarzschild like solutions in $f(R)$. For that reason we can fix $\bar{R} = 0$ along side Eq.~(\ref{Constraint on f(R) due to zero curvature}). This translates to $f(0)=0$. Under these conditions Eq.~(\ref{massive scalar eqn 1}) becomes
\begin{equation}
\label{sch:massive scalar eqn 1}
3f^{\prime\prime}(0)\bar{\Box}\delta R-(1+f^{\prime}(0))\delta R  = 0.
\end{equation}
This equation is a source-free massive scalar equation, with $\delta R$ identified as the scalar field and the mass-squared being $m^2 \equiv \frac{1+f'}{3f''}$. Henceforth, whenever there is no scope for ambiguity, we will omit the $0$ from the argument value of $f$ and all its derivatives.  This massive scalar longitudinal mode \cite{Kausar 2016,Corda 2007} is absent in GR.

Now the perturbed Einstein equation takes the following form:
\begin{equation}
\delta G_{\mu\nu} = \kappa^2\delta T^{eff}_{\mu\nu}\,,
\end{equation}
accompanied by the simplified
\begin{equation}
\label{effective EM tensor1}
\delta T^{eff}_{\mu\nu} = -\frac{\beta}{\kappa^2}\bigg[\bar{g}_{\mu\nu}\bar{\Box} -\bar{\nabla}_{\mu}\bar{\nabla}_{\nu}\bigg]\delta R\,,
\end{equation}
where $\beta = \frac{f^{\prime\prime}}{1+f^{\prime}}$. Heretofore, we will use both expressions interchangeably. The important thing to note is that there are possible $f(R)$ theories, such as $\alpha R^3$, for which $f^{\prime\prime} = 0$ (at $R=0$); in that case $\beta$ becomes zero. Another key point is that the information of a particular $f(R)$ theory enters only through the $\beta$. So, apart from $\beta$, dynamically speaking, all $f(R)$ theories are the same.

Now we use the separation of variables to isolate the angular dependence.  For this reason we write the perturbed Ricci scalar as $\delta R = \frac{\Phi (y^a)}{r}S(z^A) \equiv \Omega(y^a) S(z^A)$, where $S(z^A)$ are the scalar spherical harmonics. For radial coordinate we sometimes utilize  the tortoise coordinate $r_*$,  defined via $\frac{dr_*}{dr} = (1-\frac{2M}{r})^{-1}$. The time dependence is taken to be $e^{-i\omega t}$. After simplification, Eq.~(\ref{massive scalar eqn 1}) takes the following form:
\begin{equation}
\label{massive scalar qnm}
\frac{d^2\Phi}{dr_*^2} + (\omega^2-\tilde{V})\Phi = 0 \,,
\end{equation}
where
\begin{equation}
\tilde{V} = V_{RW} + \frac{g(r)(1+f^{\prime})}{3f^{\prime\prime}} \equiv g(r)\bigg(\frac{k^2}{r^2}+\frac{{2M}}{r^3}+\frac{1+f^{\prime}}{3f^{\prime\prime}}\bigg) \,.
\end{equation}
This result for $f(R) = \alpha R^2$ theory was discussed in Ref.~\cite{Shanki}.

It is well known that each component of a tensor behaves differently under rotation group. As a result, the effective energy momentum tensor can be separated into scalar, vector and the tensor modes~\cite{Ishibasi 2011,kodama 2000,kodama 2003}, as follows:
\begin{equation}
\label{matrix separated form of energy momentum tensor}
\delta T^{eff}_{\mu\nu} = 
\end{equation}
\[
	\left[
	\begin{array}{c|c}
	\tau_{ab} \bf{S} & r\tau_a^{(S)}\bf{S_B} \\
   \hline
   r\tau_a^{(S)}\bf{S_B} & r^2 \delta P \gamma_{AB}{\bf{S}}+r^2 \tau_T^{(S)}{\bf S}_{AB} \\ 	
	\end{array}
	\right]+
\]
\[
\left[
	\begin{array}{c|c}
	 0 & r\tau_a^{(V)}\bf{V_B} \\
   \hline
   r\tau_a^{(V)}\bf{V_B} & r^2 \tau_T^{(V)}{\bf V}_{AB} \\	
	\end{array}
	\right]+
    \]
    \[
     \left[
	\begin{array}{c|c}
	 0 & 0 \\
   \hline
   0 & r^2 \Theta_T{\bf T}_{AB} \\ 	
	\end{array}
	\right]  \,,
   \]
where $\bf{S}, \bf{V_A}$ and $\bf{T_{AB}}$ are the pure scalar, vector and tensor spherical harmonics, respectively. Rest of the tensors are defined from the pure spherical harmonic tensors. Further details, including the expression for $\Theta_T$ can be found in  Ref.~\cite{Ishibasi 2011,kodama 2000,kodama 2003} and in appendix~\ref{Spherical Harmonics}.

We find the various components of the energy momentum tensor from  Eq.~(\ref{matrix separated form of energy momentum tensor}) to be
\begin{equation}
\begin{split}
\tau_{ab} &= -\frac{f^{\prime\prime}}{\kappa^2(1+f^{\prime})}\bigg[\bar{g}_{ab}(\tilde{\Box}\Omega -\frac{k^2\Omega}{r^2}+\frac{2}{r}D^arD_a\Omega)-D_aD_b\Omega\bigg],\\
\tau_a^{(S)} &= -\frac{f^{\prime\prime}k}{\kappa^2(1+f^{\prime})}\bar{D}_a\bigg(\frac{\Omega}{r}\bigg),\\
\tau_T^{(S)} &= \frac{f^{\prime\prime}k^2}{\kappa^2(1+f^{\prime})}\bigg(\frac{\Omega}{r^2}\bigg),\\
\delta P &= -\frac{f^{\prime\prime}}{\kappa^2(1+f^{\prime})}\bigg(\tilde{\Box}-\frac{k^2}{2r^2}+\frac{1}{r}\bar{D}^ar\bar{D}_a\bigg)\Omega,
\end{split}
\end{equation}
where $\delta R = \Omega(x^a) S(z^A)$.

The scalar and vector master equations get modified as follows:
\begin{widetext}
\begin{eqnarray}
\label{qnm eqn 1}
\frac{d^2\Phi_S}{dr_*^2}+ (\omega^2-V_S)\Phi_S &=& S_S^{eff}\,,\label{qnm eqn 1a}
\\
\frac{d^2\Phi_V}{dr_*^2}+ (\omega^2-V_V)\Phi_V &=& 0 \,,\label{qnm eqn 1b}
\end{eqnarray}
\end{widetext}
where
\begin{widetext}
\begin{eqnarray}
\label{qnm eqn 1c}
S_S^{eff} &=& -\frac{\beta}{H^2 r^4}  \bigg[\Omega \bigg\{2 (2 M-r) \bigg(H r^2 \bigg(-H+k^2-2\bigg)+2 H M r+12 M^2\bigg)\nonumber \\
&&+2 H r^5 \omega ^2+P_1 r^2 (r-2 M)\bigg\}\nonumber \\
&&+2 r (r-2 M) \bigg\{2 \bigg(-(H-6) M r+Hr^2-12 M^2\bigg) \Omega ' \nonumber \\
&&+H r^2 (r-2 M) \Omega ''\bigg\}\bigg]\,,
\end{eqnarray}
\end{widetext}
and
\begin{equation}
\begin{split}
H &\equiv k^2 - 2 +\frac{6M}{r},\\
P_1 &\equiv -\frac{48M^2}{r^2} + \frac{4M}{r}(8-k^2) + 2k^2(k^2-2).
\end{split}
\end{equation}
It should be noted that as $\beta \to 0$ these equations, appropriately, reduce to their GR counterparts even if the theory is $f(R)$ with $\beta =0$ .

We also find that the tensor mode gives rise to the following equation~\cite{kodama 2003},
\begin{equation}
\tilde{\Box}\Phi_T -\frac{V_T}{g}\Phi_T = 0,
\end{equation}
where $\tilde{\Box}$ represents the $(r_*-t)$ part of the d'Alembertian operator and 
\begin{equation}
V_T = \frac{g}{r^2}\big[\lambda_L-2+\frac{2M}{r}\big]\,.
\end{equation}
Above, $\lambda_L$ is the eigenvalue of the Lichnerowicz operator~\citep{kodama 2003}.
Therefore we see that the tensor and the vector modes do not undergo any modification. This is understandable because the effective energy-momentum tensor arises from the different combinations of the derivatives of the Ricci scalar. It has been shown in Ref.~\cite{cai 2013} that the tensor harmonic functions are identically zero in transverse-traceless (TT) gauge. For this reason we will not pursue the study of the tensor mode any further.

Since the background in our studies is a $R=0$ solution, 
its metric does not have any extra hairs. However, we notice from Eq.~(\ref{qnm eqn 1}) that even though the background black hole metric has no additional hair
and is similar to the black holes in GR, at the level of perturbations the Zerilli mode gets modified from GR due to an extra source term. This source term originates because at the level of perturbations the extra massive mode that was absent in the background now gets excited. The excited massive mode ``generates" hair via $\beta$. This tells us that the perturbations will therefore get modified from GR. As a result, we can expect that the detection of QNMs through gravitational wave observations will, in principle, provide us with the opportunity to distinguish between GR and $f(R)$ theories of gravity.

\section{solution for qnm} \label{solution for qnm}

\subsection{Solution in radiation zone}

As we can see from the structure of the equations, the scalar equation gets modified by a source term. Therefore, it is understandable that the solution of that equation will have two parts: (a) One of these will be for the homogeneous part of the equation; this is identical to the GR solution; (b) The second will be the solution of the inhomogeneous part. Solution of the vector equation will be identical to the GR solution. Therefore, we can write
\begin{equation}
\label{structure of QNM}
\Phi_S = \Phi_S^{GR} + \Phi_S^{IH},\,\,\,\,\Phi_V = \Phi_V^{GR}.
\end{equation}
This implies that from the observational perspective the nonzero inhomogeneous part will be the signature of any deviation from GR. 

If we take the $r\rightarrow \infty$ limit, then we find that
\begin{equation}
S_S^{eff} \bigg|_{r\rightarrow \infty} = -\frac{2\beta}{H}  \bigg[\omega ^2 \Phi+\Phi ''\bigg],
\end{equation}
where $H = k^2-2$. However, due to Eq.~(\ref{massive scalar qnm}),
\begin{equation}
\label{massive qnm at infinity}
\frac{d^2\Phi}{dr^{*2}} = \left(\frac{1}{3\beta}-\omega^2 \right)\Phi.
\end{equation}
The source term becomes
\begin{equation}
S_S^{eff} \bigg|_{r\rightarrow \infty} = -\frac{2}{3H}\Phi.
\end{equation}
Equation~(\ref{massive qnm at infinity}) has solution of the form $e^{i\chi r_*}$, with
\begin{equation}
\chi^2 = \omega^2-\frac{1}{3\beta}.
\end{equation}
From Eq.~(\ref{qnm eqn 1}) we find,
\begin{equation}
\label{final QNM form}
\Phi_S \bigg|_{r\rightarrow \infty}= e^{i\omega r_*}-\frac{2\beta}{H}e^{i\chi r_*}.
\end{equation}
This gives the form of the inhomogeneous term arising due to the $f(R)$ theory:
\begin{equation}
\Phi^{IH}_S = -\frac{2\beta}{H}e^{i\chi r_*}.
\end{equation}
Therefore, one can see that there is a
deviation from GR in this $f(R)$ theory, howsoever tiny its effect might be on observables.

In Eq.~(\ref{final QNM form}) we derived the asymptotic deviation of the $f(R)$ QNMs from GR. This provides the opportunity to constrain $f(R)$ theories by using it. In principle, there exists the possibility of bounding $\beta$ by observationally tracking how much energy is lost by QNMs and if that deviates from the prediction of GR.
Solar system tests have already constrained $f(R)$ theories (details can be found in appendix~\ref{constraints on f(R)}). In comparison, the results found in the current work lend themselves to possible tests using 
GW observations.
The inhomogeneous term arises due to the existence of the massive polarization mode, as seen from Eqs.~(\ref{sch:massive scalar eqn 1}) and (\ref{effective EM tensor1}). This  occurs because the scalar mode and the massive mode are coupled, as was shown in Eq.~(\ref{qnm eqn 1}). Through this inhomogeneous term, energy transfer can occur between the massive mode and the scalar mode. Therefore, tests for non-GR polarizations and energy content in each polarization may, in principle, constrain these $f(R)$ theories.


Another observational aspect of these findings is the breaking of iso-spectrality. It has been shown in Ref.~\cite{Chandra}
that the scalar and vector modes have identical spectra in GR. But now due to the source term it becomes evident that in general $f(R)$ theories this iso-spectrality will not be satisfied. The homogeneous equation will have an iso-spectral solution to the vector mode but the inhomogeneous part will not be iso-spectral. This point has been discussed in Sec.~\ref{Iso-spectrality}. Ref.~\cite{Shanki} claimed that the breaking of iso-spectrality will be there for all $f(R)$ theories. We prove above that in general the iso-spectrality will be violated if the theory has nonzero $\beta$. The details of the breakdown of the iso-spectrality has been discussed in~\ref{Iso-spectrality}.

\subsection{Broken Iso-spectrality}\label{Iso-spectrality}

Breaking of iso-spectrality between the scalar and vector modes in the context of $\alpha R^2$ theory has already been discussed in Ref.~\cite{Shanki}. Even though they claim it to be a universal phenomenon for $f(R)$ theories, we showed above that this is not true unless
$\beta \neq 0$. For this reason we will investigate for the first time in detail exactly how iso-spectrality breaks down when $\beta = 0$.

Iso-spectrality in GR was first discovered by Chandrasekhar \cite{Chandra}. We will follow his methods for this investigation. Say, $Z_{1,2}$ satisfy the following equations w.r.t. $r_*$:
\begin{equation}
\label{Z equations}
\begin{split}
&\frac{d^2Z_1}{dr_*^2} + \omega^2 Z_1 = V_1 Z_1,\\
&\frac{d^2Z_2}{dr_*^2} + \omega^2 Z_2 = V_2 Z_2,
\end{split}
\end{equation}
where $V_1$ and $V_2$ are the corresponding potentials. Using the ansatz that $Z_1 = p Z_2 + qZ_2'$ and taking its derivative twice it is possible to show
\begin{equation}
Z_1'' = [p''+(p+2q')(V_2-\omega^2)+qV_2']Z_2 + [p'+q(V_2-\omega^2)+p'+q'']Z_2',
\end{equation}
where the prime denotes derivative w.r.t. $r_*$. Comparing it with Eq.~(\ref{Z equations}) Chandrasekhar showed that
\begin{equation}
\begin{split}
&q(V_1-V_2) = 2p'+q'',\\
&p(V_1-V_2) = p'' + 2q'(V_2-\omega^2) + qV_2'.
\end{split}
\end{equation}
Now, finding a solution for $p$ and $q$ establishes a relation between $Z_1$ and $Z_2$. For the QNMs of the Schwarzschild solution in GR it is possible to identify, $Z_1 = \Phi_S, Z_2 = \Phi_V, V_1 = V_S$ and $V_2 = V_V$. It is also possible to find $p$ and $q$. As a result, one can show that
\begin{equation}
\Phi_{S/V} = \frac{1}{\frac{-\lambda^2(\lambda + 1)^2}{9M^2}-\omega^2}\left(\mp W\Phi_{V/S} \, + \frac{d \Phi_{V/S}}{dr_*}\right),\,\,\,\,\,\,\,\,\,\,\,\,\,\,\,\,W = \frac{6M(2M-r)}{r^2(6M+2\lambda r)}-\frac{\lambda(\lambda + 1)}{3M},
\end{equation}
where $\lambda = \frac{(l-1)(l+2)}{2}$. Due to this relationship it becomes evident that if $\Phi_S$ depends on the radial coordinate as $e^{i\omega r_*}$ then so does $\Phi_V$. Therefore, they share the same spectrum \cite{Berti 2009}.

By inspecting Eq.~(\ref{Z equations}) it is clear that the reason for such a simplification is its homogeneous nature. If we modify Eq.~(\ref{Z equations}) with a source term, as in
\begin{equation}
\label{Z equations with source}
\begin{split}
&\frac{d^2Z_1}{dr_*^2} + \omega^2 Z_1 = V_1 Z_1 + S,\\
&\frac{d^2Z_2}{dr_*^2} + \omega^2 Z_2 = V_2 Z_2,
\end{split}
\end{equation}
then an ansatz of the previous form, $Z_1 = p Z_2 + qZ_2'$,  again leads to
\begin{equation}
Z_1'' = [p''+(p+2q')(V_2-\omega^2)+qV_2']Z_2 + [p'+q(V_2-\omega^2)+p'+q'']Z_2'.
\end{equation}
But comparison with Eq.~(\ref{Z equations with source}) does not lead to equations of $p,q$ that are independent of $Z_1$. So, in the sourced case it is not possible to 
have
$Z_1 = p Z_2 + qZ_2'$ as a solution. This eventually results in the violation of iso-spectrality.

One key point is worth mentioning. It is always possible to separate $Z_1$ as $Z_1 = Z_1^{GR} + Z_1^{IH}$, where
\begin{equation}
\frac{d^2Z_1^{GR}}{dr_*^2} + \omega^2 Z_1^{GR} = V_1 Z_1^{GR}
\end{equation}
and
\begin{equation}
\frac{d^2Z_1^{IH}}{dr_*^2} + \omega^2 Z_1^{IH} = V_1 Z_1^{IH} + S.
\end{equation}
Therefore, it is possible to have $Z_1^{GR} = pZ_2+qZ_2'$ and, consequently, there will exist an iso-spectral solution. Hence, $\Phi_S$ in $f(R)$ theories will have a part that is iso-spectral to $\Phi_V$ and another part that arises solely due to the source term that does not obey iso-spectrality.

\subsection{Massive mode\label{Massive mode}}

As already shown above in Eq.~(\ref{massive scalar eqn 1}) there exists a massive mode in $f(R)$ theories. The equation for this scalar mode is exactly equivalent to the equation of a scalar field around a Schwarzschild black hole in GR. The QNM of a massive scalar field around a Schwarzschild black hole in GR has been studied extensively~\cite{massive scalar field 1,Ohashi 2004}. Because of the similarity between the two problems, the results for the QNM frequency of the massive mode relevant for our current work will be exactly the same as the ones found in those works. Naturally, the QNM frequency values found in those works depend on the mass of the scalar field. In our current work that mass depends on the specific $f(R)$ theory chosen. Even though there are stringent restrictions on the value of this mass from solar system observations (see appendix ~\ref{constraints on f(R)}), it is possible that independent restriction may be found from GW observations. Current constraints on the mass imply that if an $f(R)$
theory is the correct theory of gravity then, $m\geq 1.25 \times 10^4 \,\,{\rm m}^{-1}$.

One key point is worth mentioning here. Usually $\omega$ is a complex number, which implies that $\chi$ should be such a number as well. Consequently, the QNMs are damped. On top of that, the large value of the mass of the massive mode will give rise to a strong damping pattern. As a result, the contribution of the inhomogeneous part in Eq.~(\ref{final QNM form}) will be very less. But massive scalar fields in Schwarzschild backgrounds can have purely real $\omega$ \cite{Konoplya 2005}. This originates mainly from the sub-dominant asymptotic contribution arising due to the irregular singularity at infinity \cite{Konoplya 2005}. These modes are called quasi-resonance frequency modes \cite{Ohashi 2004}. The requirement for the existence of such modes is $\omega_{QRM}< m$ \cite{Konoplya 2005}, where $m$ is the mass of the massive mode. Further numerical study revealed that mass has a crucial role in the damping of that mode. They showed that the greater the mass the lesser is the damping rate. Therefore, purely real modes originate corresponding to the non-damping oscillation. They also showed that for a given mass of the field a certain number of lower overtones disappear. However, this disappearance happens only for the lower overtones, while the remaining overtones are still damped. Hence, owing to the non-zero mass of the massive mode the inhomogeneous term will contain some quasi-resonance contribution that will not decay as fast as the other damped overtones.

\section{Connecting with  Scalar-Tensor theories}

It has been shown in several works that there is an equivalence between $f(R)$ theories and Brans-Dicke scalar-tensor theories (STT) \cite{Nojiri 2011, Sotiriou_2010, Felice 2010, Capozziello 1997,  Catena 2007, chiba 2013}. In this section we investigate what implication can be found from such equivalence for a perturbed Schwarzschild black hole. For this reason first we will discuss the perturbation equations in STT around a Schwarzschild black hole in Einstein frame. Then we discuss the connection between the  QNMs in $f(R)$ theory and STT.

The Einstein equations in STT can be written as,
\begin{equation}
\label{STT Einstein eqn}
 G_{\mu\nu} = R_{\mu\nu} -\frac{1}{2} R g_{\mu\nu}= T_{\mu\nu}^{STT},
\end{equation}

where,
\begin{equation}
    T_{\mu\nu}^{STT} =  \kappa^2[\nabla_{\mu}\phi\nabla_{\nu}\phi - g_{\mu\nu}(\frac{1}{2} g^{\alpha\beta}\nabla_{\alpha}\phi\nabla_{\beta}\phi + V(\phi))],
\end{equation}

and the scalar field $\phi$ satisfies,
\begin{equation}
\label{STT phi eqn}
    \Box \phi -V'(\phi) = 0\,.
\end{equation}
In this work we focus on the perturbation of a hairless Schwarzschild BH. Therefore, the LHS of Eq.~(\ref{STT Einstein eqn}) is zero for the background. This implies that for the background one has $\phi = \phi_C$, where $\phi_C$ is some constant value of the scalar field $\phi$ that satisfies $V(\phi_C) = 0$.

The perturbed part of Eq.~(\ref{STT Einstein eqn}) takes the following form,

\begin{equation}
    \delta G_{\mu\nu} = \delta T_{\mu\nu}^{STT} = -\kappa^2\bar{g}_{\mu\nu}V'(\phi)\Big|_{\phi=\phi_C} \delta \phi\,,
\end{equation}
where $\delta \phi$ is the perturbation in scalar field $\phi$.

For simplicity we will write $V'(\phi)|_{\phi=\phi_C}$ as $V'(\phi_C)$. To isolate the dynamic part of the perturbation we separate $\delta\phi$ in orbit space and angular space, as has been described earlier. For that purpose we write $\delta \phi = \frac{\Phi^{STT} (y^a)}{r}S(z^A) \equiv \Omega^{STT}(y^a) S(z^A)$, where $S(z^A)$ are the scalar spherical harmonics. For radial coordinate we use the tortoise coordinate $(r_*)$. The time dependence is taken to be $e^{-i\omega t}$. After simplification, Eq.~(\ref{STT phi eqn}) takes the following form:
\begin{equation}
\label{STT massive scalar qnm}
\frac{d^2\Phi^{STT}}{dr_*^2} + (\omega^2-\tilde{V})\Phi^{STT} = 0 \,,
\end{equation}
where
\begin{equation}
\tilde{V} = g(r)\bigg(\frac{k^2}{r^2}+\frac{2M}{r^3}+ V''(\phi_C)\bigg) \,.
\end{equation}

The scalar and vector master equations get modified as follows:
\begin{widetext}
\begin{eqnarray}
\label{STT qnm eqn 1}
\frac{d^2\Phi_S}{dr_*^2}+ (\omega^2-V_S)\Phi_S &=& S_S^{STT} = -2g(r)\frac{\kappa^2}{H} V'(\phi_C)\Phi^{STT}\,,\label{STT qnm eqn 1a}
\\
\frac{d^2\Phi_V}{dr_*^2}+ (\omega^2-V_V)\Phi_V &=& 0 \,,\label{STT qnm eqn 1b}
\end{eqnarray}
\end{widetext}
where
\begin{equation}
H \equiv k^2 - 2 +\frac{6M}{r}.
\end{equation}
Details of the calculation can be found in the appendix~\ref{effective source}.

\subsection{Radiation zone solution and broken Iso-spectrality}

As we can see from the structure of the equations, the scalar equation gets modified by a source term. Therefore, it is understandable that the solution of that equation will have two parts: (a) One of these will be for the homogeneous part of the equation, and will be identical to the GR solution; (b) The second part will be the solution of the inhomogeneous part. The solution of the vector equation will be the same as the corresponding case in GR. Therefore, we can write
\begin{equation}
\label{STT structure of QNM}
\Phi_S = \Phi_S^{GR} + \Phi_S^{IH},\,\,\,\,\Phi_V = \Phi_V^{GR}.
\end{equation}
From the observational perspective this implies that the nonzero inhomogeneous part is the signature of deviation from GR. 
If we take the $r\rightarrow \infty$ limit then we find that the source term becomes
\begin{equation}
S_S^{STT} \bigg|_{r\rightarrow \infty} = -\frac{2\kappa^2 V'(\phi_C)}{H}\Phi^{STT},
\end{equation}
where $H = k^2-2$.

Asymptotically Eq.~(\ref{STT massive scalar qnm}) has solution of the form $e^{i\chi r_*}$ with
\begin{equation}
\chi^2 = \omega^2-V''(\phi_C).
\end{equation}
Hence from Eq.~(\ref{qnm eqn 1}) we find,
\begin{equation}
\label{STT final QNM form}
\Phi_S \bigg|_{r\rightarrow \infty}= e^{i\omega r_*}-\frac{2\kappa^2 V'(\phi_C)}{H V''(\phi_C)}e^{i\chi r_*}.
\end{equation}
Therefore, the inhomogeneous term arising in the STT:
\begin{equation}
\Phi^{IH}_S = -\frac{2\kappa^2 V'(\phi_C)}{H V''(\phi_C)}e^{i\chi r_*}.
\end{equation}
Therefore, we find that there is a deviation from GR in STT, as is manifest when comparing with Eq.~(\ref{structure of QNM}), howsoever tiny the effect of this term might be on observables. Therefore, in STT also the iso-spectrality is broken as long as $V'(\phi_C) \neq 0$.

\subsection{Equivalence with $f(R)$}

In this section we will discuss the correspondence between the QNMs of a Schwarzschild BH in STT and those of such a BH in $f(R)$ theory. Let us begin by considering the $f(R)$ gravity model (known as Jordan frame) that has a Lagrangian of the following form:
\begin{equation}
\begin{split}
S_J &= \frac{1}{2\kappa^2}\int d^4x~\sqrt{-g}~f(R),\\
&= \int d^4x~\sqrt{-g}~(\frac{f'(R) R}{2\kappa^2} - \frac{f'(R) R - f(R)}{2\kappa^2}),
\end{split}
\end{equation} 
where $f(R)$ is the same as that used in Eq.~(\ref{action}); we will retain this definition until Eq.~(\ref{EOM in Einstein frame}) below.
Conformal transformation of the Jordan frame metric yields the new metric $\boldsymbol{g}_{\mu\nu} = \Omega^2 g_{\mu\nu}$, where $\Omega$ is the conformal factor. Under the identification 
\begin{equation}
\label{frame mapping}
    \Omega^2 = f'(R),\,\,\,\,\,\,\,\,\,\,\kappa \phi = \sqrt{\frac{3}{2}}{\rm{ln}}f'(R) = \sqrt{\frac{3}{2}}{\rm{ln}}\Omega^2\,,
\end{equation}
the action reduces to
\begin{equation}
    S_E = \int d^4x \sqrt{-\boldsymbol{g}}\bigg(\frac{\boldsymbol{R}}{2\kappa^2} - \frac{1}{2}\boldsymbol{g}^{\mu\nu}\nabla_{\mu}\phi\nabla_{\nu}\phi - V(\phi)\bigg)\,,
\end{equation}
which is known as the Einstein-frame action; here, the potential $V(\phi)$  can be written as
\begin{equation}
   V(\phi) = \frac{f'(\phi) R(\phi) - f(\phi)}{2\kappa^2 f'(\phi)^2}\,.
\end{equation}
 In our convention, all the metric dependent quantities that originate from a conformal transformation of the Jordan frame are boldfaced. The field equation for $\boldsymbol{g}_{\mu\nu}$ can be found by varying the Einstein frame action, and is given by \cite{Sotiriou_2010, Sumanta 2016}
 \begin{equation}
 \label{EOM in Einstein frame}
     \begin{split}
         \boldsymbol{G}_{\mu\nu} &\equiv \boldsymbol{R}_{\mu\nu} - \frac{1}{2}\boldsymbol{R}\boldsymbol{g}_{\mu\nu}\\
         &= \kappa^2 \bigg[\nabla_{\mu}\phi\nabla_{\nu}\phi-\boldsymbol{g}_{\mu\nu}\big(\frac{1}{2}\boldsymbol{g}^{\alpha\beta}\nabla_{\alpha}\phi\nabla_{\beta}\phi + V(\phi)\big)\bigg]
     \end{split}\,.
 \end{equation}
In the current work our objective is to compare the QNMs in these two theories. The background metrics are taken to be identical. In general, however, it is not possible to have identical background in both the theories that are related to each other through Eq. (\ref{frame mapping}). We now return to the usage $f(R) \rightarrow R + f(R)$, as has been described in Sec. \ref{General Result}. For Schwarzschild solution in $f(R)$ we find $\Omega^2 = 1 +f'(0)$ from Eq. (\ref{frame mapping}). For both of the theories to have the same metric as conformally related solutions we need $\Omega^2 =1$, in case of Swarzschild solution it is achievable only if $f'(0) = 0$. This is not true in general. Hence, in general, a Schwarzschild solution in $f(R)$ theory does not conformally transform to a Schwarzschild solution in STT, even though it is equivalent to some solution in STT. Therefore, QNMs in $f(R)$ theory do not conformally transform to QNMs of the same background in STT in general.

This property is not limited to $R=0$ solutions alone. For a general background, one has $\Omega^2 = 1+f'(R)$, and the two conformally related theories can have identical metric solutions if and only if $f'(R) = 0$. If this condition is not satisfied then even though the corresponding theories are conformally equivalent to each other, they can not have an identical metric solution that are conformally related.

As we are discussing QNMs of Schwarzschild BHs, we will focus on the case  $\Omega^2 = 1+f'(0)$. When $f'(0) = 0$ a Schwarzschild solution in $f(R)$ theory corresponds to a Schwarzschild solution in the corresponding STT. For this reason we will focus on this particular category. In this case $\Omega^2 = 1$ for the background. Therefore, for perturbation we can write
\begin{equation}
\label{perturbed equivalence}
    \begin{split}
        \kappa (\bar{\phi}  + \delta\phi) &= \sqrt{\frac{3}{2}}{\rm{ln}}(1 + f''(0)\delta R)\\
        &= \sqrt{\frac{3}{2}}f''(0) \delta R = \sqrt{\frac{3}{2}}f''(0)\frac{\Phi(y^a)}{r}S(Z^A),
    \end{split} 
\end{equation}
where $\delta\phi$ and $\delta R$ are perturbed scalar field in STT and perturbed Ricci scalar in $f(R)$ theory, respectively. If we separate out the angular dependence as $\delta\phi = \Phi^{STT}(y^a)S(Z^A)/r$, then we can identify
 \begin{equation}
     \Phi^{STT} = \sqrt{\frac{3}{2\kappa^2}}f''(0)\Phi(y^a).
 \end{equation}
From Eq.~(\ref{perturbed equivalence}), we see that in STT the background scalar field, $\bar{\phi} = 0$. Similarly in $f(R)$ theory extra scalar mode is not present. Therefore, the extra scalar mode gets excited in the presence of a perturbation; otherwise it remains at zero. Vector QNMs are identical to GR in both the theories. Therefore, the vector mode in $f(R)$ theory is equivalent to the vector mode in STT; in fact, they are identical.

As has already been discussed, the scalar QNM in both the theories have a homogeneous part that is iso-spectral to the vector QNM and an inhomogeneous part that is not iso-spectral to the vector QNM. The homogeneous part in both the theories is identically equal to the GR QNM. Hence, the homogeneous part of the scalar mode in $f(R)$ theory is equivalent to the homogeneous part of the scalar mode in STT: they are identical.

The difference arises in the inhomogeneous part. This difference can be understood by computing $\delta V(\phi)$,
\begin{equation}
    \delta V(\phi) = \frac{\bar{R}f''(\bar{R}) \delta R}{2\kappa^2 (1+f'(\bar{R}))^2} -\frac{(1+f'(\bar{R}))\bar{R}-(\bar{R}+f(\bar{R}))}{\kappa^2 (1+f'(\bar{R}))^3}f''(\bar{R})\delta R\,,
\end{equation}
where $\delta R$ can be represented in terms of $\delta \phi$ from Eq.~(\ref{perturbed equivalence}). For $\bar{R} = 0$ this gives $\frac{dV(\phi)}{d\phi} = 0$. Hence, the source term in the Eq.~(\ref{STT qnm eqn 1a}) vanishes for STT, leading to a vanishing inhomogeneous solution. The Regge-Wheeler and Zerilli mode in these STTs are identical to GR.

This can be understood further with an example, namely that of the $f(R) =  \alpha R^2$ theory. The conformal transformation can be expressed as follows:
\begin{equation}
    \Omega^2 = 1 + 2\alpha R,\,\,\,\,\,\,\,\,\,\,\kappa \phi = \sqrt{\frac{3}{2}}{\rm{ln}}(1 + 2\alpha R)\,.
\end{equation}
The potential of the corresponding STT can be written as
\begin{equation}
\label{r^2 potential}
   V(\phi) = \frac{ (1-e^{-\sqrt{2}\kappa\phi/\sqrt{3}})^2}{8\alpha \kappa^2}\,.
\end{equation}
A Schwarzschild solution for this theory has $\phi_C = 0$. Using it in Eq.~(\ref{r^2 potential}) we find $\frac{d V(\phi)}{d\phi}\Big|_{\phi = 0} = 0$. Hence, there are no source terms in Eq. (\ref{STT qnm eqn 1a}), thereby making it identical to GR. On the other hand, if we take $f(R) = \alpha R^2$ in Eq.~(\ref{qnm eqn 1a}), we find that $\beta = 2\alpha$, which leads to a non-vanishing source term. Therefore, the QNMs in that theory are different from those in GR, and isospectrality is violated in this example of $f(R)$. But in the corresponding STT the Zerilli QNM is exactly like GR and the iso-spectrality stays unbroken.

This result indicates some interesting features of the correspondence between $f(R)$ theory and STT. We have shown explicitly that the iso-spectrality breaking is not present in the QNMs after one transforms to the Einstein frame using a conformal transformation. A particular metric solution of an $f(R)$ theory gets mapped to some metric solution of the conformally related scalar-tensor theory. This by itself does not, however, imply that the two spacetime solutions are identical; therefore, it is not necessary for them to share all of the same characteristics. Indeed, in general, a particular solution in $f(R)$ will not get mapped to exactly the same solution in the conformally related scalar-tensor theory. This is often the case when the conformal factor is not unity. In that case there is no obvious reason for their QNMs to share similar properties. 
What we have shown here is that when an $f(R)$ there and an STT both have the Schwarzschild metric as solution and conformal factor relating these respective solutions is unity, then isopectrality is not broken for the Schwarzschild QNMs in STT even though it is broken for the Schwarzschild QNMs in $f(R)$.

\section{Discussion} \label{conclusion and discussion}

In this work we have focused on the QNMs of a Schwarzschild black hole in a general $f(R)$ theory that allows the $R=0$ solution. We came to the conclusion that in general $f(R)$ theories the iso-spectrality between the scalar and the vector mode will not be valid. Breaking of iso-spectrality was discussed in the work \cite{Shanki} for the specific case of $\alpha R^2$ theory. They claimed this result will be true for all other theories also. We found that this claim is not entirely true. In general, there will be breaking of isospectrality, but it is possible to have theories where it is not true, such as when $f(R) = \alpha R^3$. Indeed, it is possible to conclude that the isospectrality will be violated if $f''/(1+f') \neq 0$. 

Secondly, we discussed that the structure of QNMs gets modified at the asymptote. This modification arises as an inhomogeneous term alongside the GR contribution. Therefore, not only the iso-spectrality will be violated but also the QNM structure will get modified. But for $f''/(1+f') = 0$, this contribution along with the breaking of the isospectrality will be absent, resulting in zero deviation from GR in first order perturbation. This does not imply that the ringdown spectrum will be exactly like GR. If we take second order perturbation then we find,
\begin{equation}
    \begin{split}
        \delta^2 G_{\mu\nu} &= \frac{\mathcal{S_{\mu\nu}}}{1+f'}\\
     \mathcal{S}_{\mu\nu} &= f'' \delta R \delta R_{\mu\nu} + f'''\bar{g}_{\mu\nu}\delta R \bar{\Box} \delta R + f'' \delta g_{\mu\nu}\Box \delta R + f'' \bar{g}_{\mu\nu} \delta (\bar{\Box} \delta R) + f'''\frac{\bar{R}}{4}\bar{g}_{\mu\nu}\delta R^2
         - f'' \frac{\bar{g}_{\mu\nu}}{4}\delta R^2\\
         &+ f'' \frac{\bar{R}}{4}\delta g_{\mu\nu}\delta R + f''\frac{\bar{R}}{4}\bar{g}_{\mu\nu}\delta^2R - f''\delta(\nabla_{\mu}\nabla_{\nu}\delta R) - f'''\delta R\nabla_{\mu}\nabla_{\nu}\delta R  +(f-Rf')\frac{\delta^2g_{\mu\nu}}{2}.
    \end{split}
\end{equation}

For $\bar{R} = 0$ solution $\mathcal{S}$ reduces to,
\begin{equation}
  \begin{split}
      \mathcal{S}_{\mu\nu} &= f'' \delta R \delta R_{\mu\nu} + f'''g_{\mu\nu}\delta R \Box \delta R + f'' \delta g_{\mu\nu}\Box \delta R + f'' g_{\mu\nu} \delta (\Box \delta R)
         - f'' \frac{g_{\mu\nu}}{4}\delta R^2\\
         & - f''\delta(\nabla_{\mu}\nabla_{\nu}\delta R) - f'''\delta R\nabla_{\mu}\nabla_{\nu}\delta R. 
  \end{split}
\end{equation}

As it depends on $f'''$, the ringdown mode will contain information of the theory but the first order QNM will be exactly like GR. Therefore a detailed numerical relativity evolution will capture these deviations. As a result one can in principle constrain $\beta$ and $\frac{d^nf(R)}{dR^n}\big|_{\bar{R} = 0}$ solely from GW observations.

One key-point is worth mentioning here. In our current work we have focused only on the perturbation around a vacuum $\bar{R}=0$ solution. We have used these similar assumptions while interpreting the solar system test. Owing to this choice we have not been able to pursue the study of Chameleon screening in $f(R)$ theories \cite{Burrage 2018}.
This screening behaviour is the origin of the different behaviour of the theory at different scales. Due to the screening effect the mass of the massive scalar mode depends on the mass density of energy momentum tensor $(T_{\mu\nu})$ in the space-time. Effective mass of the massive scalar mode becomes heavier in the region with high mass density and
lighter in the low density region.
Therefore, in different environments the mass of the massive scalar mode will have different values depending on the density profile. This gives rise to the screening mechanism.  Therefore, interpreting and using the solar system constraints should be done carefully. But unfortunately due to vacuum assumption $(T_{\mu\nu} = 0)$, we have not been able to explore this sector in our current work. Hence, a rigorous analysis with non-vacuum condition is an important endeavour that may be taken up in the future.

QNMs in scalar-tensor theories have recently been calculated in Refs.~\cite{Tattersall 2018_1, Tattersall 2018_2, Kobayashi 2012, Kobayashi 2014, Apratim 2018}. It has been shown before that a class of scalar-tensor theories are equivalent to $f(R)$ theories. Tattersall et al. have also found results similar the ones noted here in the context of scalar-tensor theories. Due to the equivalence of scalar-tensor theories and $f(R)$ theories this is, of course, expected. The equivalence between Brans-Dicke with $\omega_0 =0$ and $f(R)$ theory is valid under the condition $f''(R)\neq 0$ (related to the sufficient condition for invertibility between the two theories). When $f''$  is not defined, or vanishes, the
equality $\phi = f'(R)$ and the equivalence between the two theories cannot be guaranteed, although this is not {\it a priori} excluded by $f'' = 0$~\cite{Sotiriou_2010}. Therefore, there are families of $f(R)$ that in principle are not equivalent to scalar-tensor theories. Interestingly, in our work we see that this is precisely the sector where $f(R)$ does not bring any modification w.r.t. GR, mainly due to the vanishing coupling between the GR mode and the extra massive scalar mode.

\section*{Acknowledgments}

We thank Sumanta Chakraborty, S. Shankaranarayanan and  Varun Sahni for the useful inputs. This work is supported in part by the Navajbai Ratan Tata Trust. SD would like to thank University Grants Commission (UGC), India, for financial support as senior research fellow.

\appendix
\section{Spherical Harmonics}\label{Spherical Harmonics}

Any field on a spherically symmetric background can be classified as a scalar, vector and tensor depending on their transformation under the rotation group. For this reason three kinds of spherical harmonics are used to separate out the angular dependence.

One subset of these functions are the scalar spherical harmonics. Any scalar function on a spherically symmetric spacetime is expressible as a sum of the scalar spherical harmonics, $Y_{lm}(\theta,\phi)$. We call them $S(Z^A)$.

A vector field $U_A$ can be expressed as, $U_A = V_A + D_AS$, where $S$ is the scalar spherical harmonic defined above. $V_A$ are called the vector spherical harmonics.

A tensor field $X_{AB}$ on spherically symmetric spacetime can be expressed as, 
\begin{equation}
X_{AB} = T_{AB} + 2D_{(A}V_{B)} + \hat{L}_{AB}S + \gamma_{AB}S,
\end{equation}
where $\gamma_{AB}$ is the metric on the two sphere. $V_A$ and $S$ are the vector and scalar spherical harmonics, respectively. $\hat{L}_{AB}$ is the Lichnerowicz operator defined as, $\hat{L}_{AB} = D_AD_B - \frac{1}{2}\gamma_{AB}$. $S,\,\,V_A\,\,{\rm and }\,\, T_{AB}$ are called the pure spherical harmonic functions.

They satisfy the following equations,
\begin{equation}
\begin{split}
(\hat{\Box} + k^2)S =& 0,\\
(\hat{\Box} + k^2)V_A =& 0,\\
(\hat{\Box} + k^2)T_{AB} =& 0,
\end{split}
\end{equation}
where $k^2 = l(l+1)$, and $l$ is a non-negative integer.

From the pure spherical harmonics several useful vector and tensor fields can be defined. We only mention a few among them, namely,
\begin{equation}
\begin{split}
S_A =& -\frac{1}{k}D_AS,\\
S_{AB} =& \frac{1}{k^2}D_AD_BS + \frac{1}{2}\gamma_{AB}S,\\
V_{AB} =& -\frac{1}{2k}(D_AV_B+D_BV_A).
\end{split}
\end{equation}

\section{Separation of effective energy momentum tensor}

Let us assume that the metric of a manifold is as follows,
\begin{equation}
\begin{split}
ds^2 &= g_{ab}(x)dx^adx^b + r^2(x)d\Omega^2,\\
d\Omega^2 &= \gamma_{AB}dz^Adz^B.
\end{split}
\end{equation}
In this case the Christoffel symbols take particular forms as discussed below. A hat represents connection on two sphere and $^2\Gamma^a_{bc}$ represents the connection in the orbit space. The nonzero components of the connection are as follows,

\begin{equation}
\label{conection set}
\begin{split}
\Gamma^a_{bc} =& ^2\Gamma^a_{bc},\\
\Gamma^a_{BC} =&-rD^ar\gamma_{BC},\\
\Gamma^A_{aB} =&\frac{D_ar}{r}\delta^A_B,\\
\Gamma^A_{BC} =&\hat{\Gamma}^A_{BC}.
\end{split}
\end{equation}
Using these results for connection we can calculate various covariant derivatives. Various components of double co-variant derivatives of a scalar function $F(x^a,z^A)$ can be calculated to be as follows,
\begin{equation}
\label{derivative set}
\begin{split}
\nabla_a\nabla_b F =& D_aD_bF,\\
\nabla_a\nabla_BF =& rD_a\bigg(\frac{1}{r}D_BF\bigg),\\
\nabla_A\nabla_bF =& D_AD_bF - \frac{D_br}{r}D_AF,\\ 
\nabla_A\nabla_BF =& D_AD_BF + rD^arD_aF \gamma_{AB},\\
\Box F =& \tilde{\Box}F + \frac{1}{r^2}\hat{\Box}F + \frac{2}{r}D^arD_aF.
\end{split}
\end{equation}

It has been shown that the Einstein equation gets modified in our case. As result we get an effective energy momentum tensor of the following form,
\begin{equation}
\label{effective EM tensor}
\delta T^{eff}_{\mu\nu}=-\frac{\beta}{\kappa^2}\bigg[\bar{g}_{\mu\nu}\bar{\Box} -\bar{\nabla}_{\mu}\bar{\nabla}_{\nu}\bigg]\delta R.
\end{equation}
Since $\delta R$ is a scalar function, we write it as,
\begin{equation}
\delta R = \Omega (x^a)S(z^A).
\end{equation}
Now using the expressions in [\ref{conection set}] and [\ref{derivative set}] in Eq.~[\ref{effective EM tensor}] and using the separation explained in Eq.~[\ref{matrix separated form of energy momentum tensor}] we find,
\begin{equation}
\begin{split}
\tau_{ab} &= -\frac{\beta}{\kappa^2}\bigg[\bar{g}_{ab}(\tilde{\Box}\Omega -\frac{k^2\Omega}{r^2}+\frac{2}{r}D^arD_a\Omega)-D_aD_b\Omega\bigg],\\
\tau_a^{(S)} &= -\frac{\beta}{\kappa^2} k\bar{D}_a\bigg(\frac{\Omega}{r}\bigg),\\
\tau_T^{(S)} &= \frac{\beta}{\kappa^2} k^2\bigg(\frac{\Omega}{r^2}\bigg),\\
\delta P &= -\frac{\beta}{\kappa^2}\bigg(\tilde{\Box}-\frac{k^2}{2r^2}+\frac{1}{r}\bar{D}^ar\bar{D}_a\bigg)\Omega.
\end{split}
\end{equation}
All other terms arising in Eq.~[\ref{matrix separated form of energy momentum tensor}] are zero.

For STT $\delta T_{\mu\nu} = -\bar{g}_{\mu\nu} \frac{d V}{d\phi} \delta \phi$.

\begin{equation}
    \begin{split}
        \tau_{ab} =& -\bar{g}_{ab} \frac{d V}{d\phi} \frac{\Phi^{STT}}{r}\\
\tau_a =& \tau_T^{(S)} = 0\\
r^2\delta P =& -\frac{dV}{d\phi} \frac{\Phi^{STT}}{r}.
    \end{split}
\end{equation}

\section{Effective source}
\label{effective source}

The perturbation equations with a source term around the Schwarzschild metric was calculated in \cite{Ishibasi 2011}. We take the appropriate limit i.e. $n=2$ and put electromagnetic perturbation to zero. After doing that the result found is,
\begin{equation}
\begin{split}
S_S^{eff} = \frac{g\kappa^2}{rH}\bigg[&-HS_T-\frac{P_1 S_t}{i\omega H}-4g\frac{r S_t^{\prime}}{i\omega}-4rgS_r+\frac{P_2 r S^r_t}{i\omega H}+2r^2\frac{S^{r\prime}_t}{i\omega}+2r^2S^r_r\bigg],
\end{split}
\end{equation}
where prime denotes radial derivative.

The functions in the previous equation are dependent on the effective energy momentum tensor and the metric in the following manner,

\begin{equation}
\begin{split}
S_{ab} &= \tau_{ab},\\
S_a &= \frac{r\tau_a}{k},\\
S_T &= \frac{2r^2}{k^2}\tau_T,\\
H &= k^2 - 2 +\frac{6M}{r},\\
P_1 &= -\frac{48M^2}{r^2} + \frac{4M}{r}(8-k^2) - 2k^2(k^2-2),\\
P_2 & = \frac{24M}{r}.
\end{split}
\end{equation}

After a little calculation it can be expressed as,

\begin{widetext}
\begin{equation}
\begin{split}
S_S^{eff} =& -\frac{\beta}{H^2 r^4}  \bigg[\Omega \bigg\{2 (2 M-r) \bigg(H r^2 \bigg(-H+k^2-2\bigg)+2 H M r+12 M^2\bigg)+2 H r^5 \omega ^2+P_1 r^2 (r-2 M)\bigg\}\\
&+2 r (r-2 M) \bigg\{2 \bigg(-(H-6) M r+Hr^2-12 M^2\bigg) \Omega '+H r^2 (r-2 M) \Omega ''\bigg\}\bigg].
\end{split}
\end{equation}
\end{widetext}

For STT,

\begin{equation}
\begin{split}
S_{ab} &= \tau_{ab},\\
S_a &= 0,\\
S_T &= 0.
\end{split}
\end{equation}

\begin{equation}
    S_S^{STT} = -\frac{2  g \kappa^2}{H}\frac{d V}{d\phi} \Phi^{STT}.
\end{equation}

\section{GW propagation in Minkowski background}
We take the opportunity of the current work to discuss the propagation of GW in Minkowski spacetime for general $f(R)$ theory of gravity. To study the propagation of GW in general $f(R)$ theory, in Minkowski background we take $\bar{g}_{\mu\nu} = \eta_{\mu\nu}$.
As a result the perturbation equations become
\begin{equation}
\begin{split}
&\delta G^{eff}_{\mu\nu}=\bigg[\delta R_{\mu\nu}-\frac{1}{2}\eta_{\mu\nu}\delta R\bigg]-\frac{f(0)}{(1+f^{\prime}(0))}\frac{h_{\mu\nu}}{2},\\
&\delta T^{eff}_{\mu\nu}=-\frac{f^{\prime\prime}}{\kappa^2(1+f^{\prime}(0))}\bigg[\eta_{\mu\nu}\bar{\Box} -\partial_{\mu}\partial_{\nu}\bigg]\delta R(h).
\end{split}
\end{equation}
Taking the trace of Eq.~[\ref{perturbation eqn 1}] it can be found that,
\begin{equation}
\label{massive scalar eqn 2}
\begin{split}
3f^{\prime\prime}\partial_{\mu}\partial^{\mu}\delta R-(1+f^{\prime}(0))\delta R &= 0,\\
\Box\delta R+m^2\delta R &= 0,
\end{split}
\end{equation}
where $m^2 = -\frac{(1+f^{\prime})}{3f^{\prime\prime}}=-\frac{1}{3\beta}$. Owing to the real-valuedness of $m$, it must be the case that $\beta \leq 0$\cite{Schmidt,Berry}. This implies that $\frac{1+f'}{f''} \leq 0$.

Along the lines of Ref.~\cite{Berry} we define the infinitesimal Gauge transformation
\begin{equation}
h'_{\mu\nu} = h_{\mu\nu} -\big(\beta\delta R+ \frac{h}{2}\big)\eta_{\mu\nu}\,,
\end{equation}
and choose the Lorentz gauge condition,
\begin{equation}
\partial_{\mu}h'^{\mu\nu} = 0.
\end{equation}
Defining $\delta G^{eff}_{\mu\nu}-\kappa^2\delta T^{eff}_{\mu\nu}\equiv \delta\mathcal{G}_{\mu\nu}$, it can be shown that
\begin{equation}
\delta\mathcal{G}_{\mu\nu} = -\frac{1}{2}\Box h'_{\mu\nu}.
\end{equation}
The equation $\delta\mathcal{G}_{\mu\nu} = 0$ then implies
\begin{equation}
\Box h'_{\mu\nu} = 0.
\end{equation}
Therefore, we can see that the only difference the general $f(R)$ theory makes relative to the case studied in Ref.~\cite{Berry} is that the $a_2$ in their work gets replaced by $\beta = \frac{f''}{1+f'}$. 

\section{Constraints on $f(R)$\label{constraints on f(R)}}

In the work \cite{Berry} different methods for constraining $f(R)$ models have been discussed. We will mainly focus on the fifth force test. For this reason it is important to take the Newtonian limit. The result will be similar to that found in Ref.~\cite{Berry} by replacing $\Upsilon$ with $m$.
The metric then takes the form,
\begin{equation}
\label{Yukawa metric}
\begin{split}
ds^2 = \bigg\{1-\frac{2M}{r}\bigg[1+\frac{exp(-mr)}{3}\bigg]\bigg\}dt^2-\bigg\{1+\frac{2M}{r}\bigg[1-\frac{exp(-mr)}{3}\bigg]\bigg\}d\Sigma^2,
\end{split}
\end{equation}
where, $d\Sigma^2 = dx^2+dy^2+dz^2$.

The structure of the metric \cite{Berry} gives rise to the potential of Yukawa type, as was already discussed in \cite{Berry}. In fifth-force tests 
such potential has been studied extensively, where the considered form is,
\begin{equation}
V(r) = \frac{M}{r}\bigg[1+\alpha\, exp(-\frac{r}{\lambdabar})\bigg].
\end{equation}
From the E\"{o}t-Wash experiment \cite{Kapner 2007,Hoyle 2004} it is possible to put strong constraint, $\lambdabar = m^{-1}\leq 8 \times 10^{-5}\rm{m}$. Therefore, we are finding that for general $f(R)$ this puts up the bound $|\beta| = \big|\frac{f''}{1+f'}\big|\leq 2 \times 10^{-9} \rm{m^2}$.


\begin{references}



 \bibitem{LIGO detection 1} 
 B. P. Abbott et al. (LIGO Scientific and Virgo Collaboration)
  Phys. Rev. Lett. {\bf118}, 221101 (2017).
  
  \bibitem{LIGO detection2}
  B. P. Abbott et al. (LIGO Scientific and Virgo Collaboration)
  Phys. Rev. Lett. {\bf116}, 241103 (2016).
  
  \bibitem{LIGO detection3}
  B. P. Abbott et al. (LIGO Scientific and Virgo Collaboration)
  Phys. Rev. X {\bf6}, 041015 (2016).
  
  \bibitem{LIGO detection4}
  B. P. Abbott et al. (LIGO Scientific and Virgo Collaboration)
  Phys. Rev. Lett. {\bf116}, 061102 (2016).
  

\bibitem{First BNS observation} 
B. P. Abbott {\it et al.} (LIGO Scientific and Virgo Collaborations), 
Phys. Rev. Lett. {\bf 119}, 161101 (2017).


\bibitem{LSC constrain on EOS}
The LIGO Scientific Collaboration, the Virgo Collaboration,
arXiv:1805.11581.


\bibitem{Soumi De 2018}
Soumi De, Daniel Finstad, James M. Lattimer, Duncan A. Brown, Edo Berger, Christopher M. Biwer
arXiv:1804.08583.



\bibitem{Tianqi Zhao 2018}
Tianqi Zhao, James M. Lattimer
arXiv:1808.02858.



\bibitem{Eemeli Annala 2018}
Eemeli Annala, Tyler Gorda, Aleksi Kurkela, and Aleksi Vuorinen
Phys. Rev. Lett. 120, 172703 – Published 25 April 2018.


\bibitem{Radicel 2018}
David Radice1, Albino Perego, Francesco Zappa, and Sebastiano Bernuzzi,
The Astrophysical Journal Letters, Volume 852, (2018).




\bibitem{Abott 2017 graviton mass}
B. P. Abbott et al. (LIGO Scientific and Virgo Collaboration),
Phys. rev. Lett. 118, 221101 (2017).




\bibitem{Baker 2017}
T. Baker, E. Bellini, P. G. Ferreira, M. Lagos, J. Noller, and I. Sawicki,
Phys. Rev. Lett. {\bf 119}, 251301 (2017).


\bibitem{Ezquiaga 2017}
J. M. Ezquiaga and M. Zumalacárregui,
Phys. Rev. Lett  {\bf 119}, 251304 (2017).


\bibitem{Lucas 2016}
Lucas Lombriser and Andy Taylor,
JCAP03(2016)031


\bibitem{Lucas 2017} 
L. Lombriser, Nelson A.Lima,
Physics Letters B {\bf 765}, (2017)


\bibitem{Abott 2016}
B. P. Abbott et al. (LIGO Scientific and Virgo Collaboration)
Phys. Rev. Lett. {\bf 116}, 221101 (2016).


\bibitem{Konoplya 2016}
R.~Konoplya and A.~Zhidenko,
  Phys. Lett. B {\bf 756}, 350 (2016).



\bibitem{LISA 1}
P. Amaro-Seoane et al., 
arXiv:1702.00786.


\bibitem{LISA 2}
M. Armano, H. Audley, G. Auger, J. T. Baird, M. Bassan,
P. Binetruy, M. Born, D. Bortoluzzi, N. Brandt
et al.,
Phys.Rev. Lett. {\bf116}, 231101 (2016).


\bibitem{LISA 3}
P. A. Seoane, S. Aoudia1, S. Babak1, P. Binétruy, E. Berti, A. Bohé, C. Caprini, M. Colpi, N. J Cornish, K. Danzmann, J. F. Dufaux, J. Gair, O. Jennrich, P. Jetzer, A. Klein, R. N. Lang, A. Lobo, T. Littenberg, S. T. McWilliams, G. Nelemans, A. Petiteau, E. K Porter, B. F. Schutz, A. Sesana, R. Stebbins, T. Sumner, M. Vallisneri, S. Vitale, M. Volonteri and H. Ward,
Classical and Quantum Gravity, {\bf29}, 12, 124016, (2012).
 
 
\bibitem{LISA 4}
S. Vitale, 
Gen Relativ Gravit (2014) {\bf46}: 1730.

\bibitem{LISA 5}
S. Babak, J. Gair, A. Sesana, E. Barausse, C. F. Sopuerta, C. P. L. Berry, E. Berti, P. Amaro-Seoane, A. Petiteau, and A. Klein
Phys. Rev. D {\bf95}, 103012 (2017).


\bibitem{Will book}
C. M. Will,
\textit{Theory and Experiment in Gravitational Physics} (Cambridge, University Press, Cambridge, England, 1993).


\bibitem{will 2014}
C. M. Will,
Living Rev. Relativ. 17, 4 (2014).



\bibitem{Kormendy 1995}
Kormendy, J. and Richstone, D., Annu. Rev. Astron. Astrophys. 33, 581–624 (1995).




\bibitem{Schodel 2002}
R. Schödel, T. Ott, R. Genzel, R. Hofmann, M. Lehnert, A. Eckart, N. Mouawad, T. Alexander, M. J. Reid, R. Lenzen, M. Hartung, F. Lacombe, D. Rouan, E. Gendron, G. Rousset, A.-M. Lagrange, W. Brandner, N. Ageorges, C. Lidman, A. F. M. Moorwood, J. Spyromilio, N. Hubin and K. M. Menten,
Nature volume 419, pages 694–696 (17 October 2002).



\bibitem{Stelle 1977}
K. S. Stelle, 
Phys. Rev. D {\bf 16}, 953 (1977).

\bibitem{Stelle 1978}
K. S. Stelle,
Gen. Relativ. Gravit. {\bf 9}, 353 (1978).



\bibitem{Shahid 1990}
Shahid-Saless, B., 1990, J. Math. Phys. 31, 242.


\bibitem{Brandenberger 1992}
Brandenberger, R. H., 1992, e-print arXiv:gr-qc/9210014.



\bibitem{Brandenberger 1993}
Brandenberger, R. H., 1993, e-print arXiv:gr-qc/9302014.

\bibitem{Nojiri 2011}
Shin'ichi Nojiri, Sergei D. Odintsov.
Phys.Rept. 505 (2011).


\bibitem{Nojiri 2017}
S. Nojiri, S.D. Odintsov, V.K. Oikonomou,
Phys.Rept. 692 (2017)


\bibitem{Brandenberger et al. 1993}
Brandenberger, R. H., V. F. Mukhanov, and A. Sornborger, 1993, Phys. Rev. D 48, 1629.


\bibitem{Trodden 1993}
Trodden, M., V. F. Mukhanov, and R. H. Brandenberger, 1993, Phys. Lett. B 316, 483.


\bibitem{Starobinsky 1980}
Starobinsky, A. A., 1980, Phys. Lett. 91B, 99.






\bibitem{Hu}
W. Hu and I. Sawicki
Phys. Rev. D 76, 064004 – Published 10 September 2007.

\bibitem{Nojiri 2003}
Shin'ichi Nojiri, Sergei D. Odintsov,
Phys.Rev. D68 (2003) 123512


\bibitem{Cognola 2008}
G. Cognola, E. Elizalde, S. Nojiri, S.D. Odintsov, L. Sebastiani, S. Zerbini,
Phys.Rev. D77 (2008) 046009


\bibitem{Birrell 1982}
Birrell, N. D., and P. C. W. Davies, 1982, Quantum Fields in Curved Spacetime Cambridge University Press, Cambridge.




\bibitem{Buchbinder 1992}
Buchbinder, I. L., S. D. Odintsov, and I. L. Shapiro, 1992, Effective Actions in Quantum Gravity IOP, Bristol. 



\bibitem{Vilkovisky 1992}
Vilkovisky, G. A., 1992, Class. Quantum Grav. 9, 895.



\bibitem{Konoplya 2019}
  R.~A.~Konoplya, A.~F.~Zinhailo and Z.~Stuchlík,
  arXiv:1903.03483 [gr-qc].
  
  

\bibitem{Zinhailo 2018}
  A.~F.~Zinhailo,
  Eur.\ Phys.\ J.\ C {\bf 78}, 992 (2018)
  



\bibitem{Zhang 2011}
X. Zhang and Y. Ma,
Phys. Rev. D {\bf84}, 064040 (2011).


\bibitem{olmo 2009}
 G. J. Olmo and P. Singh,
 JCAP01 {\bf 30},(2009).




\bibitem{Regge 1957}
T. Regge and J. A. Wheeler, 
Phys. Rev. {\bf 108}, 1063 (1957).


\bibitem{Zerilli 1970}
F. J. Zerilli, 
Phys. Rev. Lett. {\bf 24}, 737 (1970).  


\bibitem{Newman 1962}
E. Newman and R. Penrose,
J. Math. Phys. (N.Y.) {\bf 3}, 566 (1962).

\bibitem{Vishu 1970}
C. V. Vishveshwara, 
Nature (London) {\bf 227}, 936 (1970).


\bibitem{Chandra} 
S. Chandrasekhar,
\textit{The Mathematical Theory of Black Holes}~(Oxford University press, New Delhi 2010).


\bibitem{Nollert 1999}
H. P. Nollert,
Classical Quantum Gravity {\bf 16}, R159 (1999).


\bibitem{Kokkotas 1999}
K. D. Kokkotas and B. G. Schmidt, 
Living Rev. Relativ. {\bf 2},2 (1999).


\bibitem{Konoplaya 2011}
R. A. Konoplaya and A. Zhidenko, 
Rev. Mod. Phys. {\bf 83}, 793 (2011).



\bibitem{Shanki}
S. Bhattacharya and S. Shankaranarayanan,
Phys. Rev. D {\bf96}, 064044 (2017).





\bibitem{Soham 2018}
S. Bhattacharyya and S. Shankaranarayanan, 
Eur. Phys. J. C (2018) 78: 737.



\bibitem{Tattersall 2018_1}
Oliver J. Tattersall and Pedro G. Ferreira
Phys. Rev. D 97, 104047 – Published 25 May 2018.



\bibitem{Tattersall 2018_2}
Oliver J. Tattersall, Pedro G. Ferreira, and Macarena Lagos
Phys. Rev. D 97, 044021 – Published 16 February 2018.




\bibitem{Woodard 2007}
R. P. Woodard,
in \textit{3rd Aegean Summer School: The Invisible Universe: Dark Matter and Dark Energy}, edited by L. Papantonopoulos (Springer, Berlin, Heidelberg, 2007), Vol. 720, pp. 430-433.


\bibitem{Myung}
Yun Soo Myung, Taeyoon Moon, and Edwin J. Son
Phys. Rev. D {\bf83}, 124009.




\bibitem{Ishibasi 2011}
A. Ishibashi and H. Kodama,
 Prog. Theor. Phys. Suppl. 189, 165 (2011).
 
\bibitem{kodama 2000}
H, Kodama, A. Ishibashi and O. Seto, 
Phys. Rev. D 62, 064022 (2000).

\bibitem{kodama 2003}
H. Kodama and A. Ishibashi, 
Prog. Theor. Phys. 110, 701 (2003).


\bibitem{Gerlach 1979}
U. H. Gerlach and U. K. Sengupta,
Phys. Rev. D {\bf19}, 2268 (1979).


\bibitem{Capozziello 2008}
S. Capozziello, C. Corda and M. F. De Laurentis,
Phys. Lett. B {\bf 669}, 255 (2008).



\bibitem{Berry}
C. P. L. Berry and J. R. Gair,
Phys. Rev. D 83, 104022 (2011).



\bibitem{Nishizawa 2009}
A. Nishizawa, A. Taruya, K. Hayama, S. Kawamura and M. Sakagami,
Phys. Rev. D {\bf 79}, 082002 (2009).



\bibitem{Kausar 2016}
H. R. Kausar, L. Philippoz and P. Jetzer
Phys. Rev. D {\bf 93}, 124071 (2016).



\bibitem{Corda 2007}
C. Corda, 
2007, J. Cosmol. Astropart. Phys. 0704, 009.




\bibitem{cai 2013}
R. G. Cai and L. M. Cao, 
Phys. Rev. D {\bf 88}, 084047 (2013).




\bibitem{Berti 2009}
E. Berti, V. Cardoso and A. O Starinets,
Class. Quantum Grav. {\bf26} 163001 (2009).



\bibitem{Martel 2005}
K. Martel and E. Poisson,
Phys. Rev. D {\bf71}, 104003 (2005).



\bibitem{Nagar 2005}
A. Nagar and L. Rezzolla,
Classical Quantum Gravity {\bf22}, R167 (2005).




\bibitem{massive scalar field 1}
L. E. Simone and C. M. Will,
Class. Quantum Grav.9 (1992).

\bibitem{Ohashi 2004}
A. Ohashi and M. A. Sakagami,
Class. Quantum Grav. 21 3973 (2004).



\bibitem{Konoplya 2005}
R. A. Konoplya and A. V. Zhidenko,
Physics Letters B, {\bf609},(2005).





\bibitem{Sotiriou_2010}
T. P. Sotiriou and V. Faraoni,
REVIEWS OF MODERN PHYSICS, {\bf82}, (2010).




\bibitem{Felice 2010}
De Felice, A., Tsujikawa, S. Living Rev. Relativ. (2010) 13: 3. https://doi.org/10.12942/lrr-2010-3


\bibitem{Capozziello 1997}
S. Capozziello, R. de Ritis, A.A. Marino, 
Class. Quant. Grav. 14, 3243–3258 (1997)




\bibitem{Catena 2007}
Riccardo Catena, Massimo Pietroni, and Luca Scarabello
Phys. Rev. D 76, 084039 – Published 31 October 2007


\bibitem{chiba 2013}
Takeshi Chiba and Masahide Yamaguchi
Journal of Cosmology and Astroparticle Physics, Volume 2013, October 2013





\bibitem{Schmidt}
H.-J. Schmidt, Astro. Nachr. 307, 339 (1986).



\bibitem{Kobayashi 2012}
Kobayashi, Tsutomu and Motohashi, Hayato and Suyama, Teruaki
Phys. Rev. D 85, 084025 (2012).



\bibitem{Apratim 2018}
Ganguly, Apratim and Gannouji, Radouane and Gonzalez-Espinoza, Manuel and Pizarro-Moya, Carlos,
Class. Quant. Grav. 35, (2018).



\bibitem{Kobayashi 2014}
Kobayashi, Tsutomu and Motohashi, Hayato and Suyama Teruaki, 
Phys. Rev. D 89, 084042 (2014)





\bibitem{Burrage 2018}
C. Burrage and J. Sakstein, Living Rev. Relativity 21, 1
(2018).





\bibitem{Kapner 2007}
D. J. Kapner, T. S. Cook, E. G. Adelberger, J. H. Gundlach,
B. R. Heckel, C. D. Hoyle, and H. E. Swanson, 
Phys. Rev. Lett. {\bf98}, 021101 (2007).



\bibitem{Hoyle 2004}
 C. D. Hoyle, D. J. Kapner, B. R. Heckel, E. G. Adelberger,
J. H. Gundlach, U. Schmidt, and H. E. Swanson,
Phys. Rev. D {\bf70}, 042004 (2004).


\bibitem{Sumanta 2016}
Chakraborty, S., SenGupta, S. Eur. Phys. J. C (2016) 76: 552. https://doi.org/10.1140/epjc/s10052-016-4394-0











\end{references}
\end{document}